\documentclass[%
 reprint,
 amsmath,amssymb,
 aps,
]{revtex4-2}
\usepackage{comment}
\usepackage{graphicx}% Include figure files
\usepackage{dcolumn}% Align table columns on decimal point
\usepackage{bm}% bold math
\usepackage{hyperref}
\usepackage{dirtytalk}
\usepackage{float}
\hypersetup{
    colorlinks=true,
    linkcolor=blue,
    filecolor=cyan,      
    urlcolor=red,
    pdftitle={Overleaf Example},
    pdfpagemode=FullScreen,
    }
\usepackage[skip=5pt, indent=10pt]{parskip}
\begin{document}

\preprint{APS/123-QED}
\title{Influence of magnetic and electric fields on universal conductance fluctuations in thin films of the Dirac semi-metal \texorpdfstring {Cd$_3$As$_2$}{Cd3As2}}% Force line breaks with \\

\author{Run Xiao}
%Lines break automatically orn be forced with \\
 %\email{}
 \affiliation{Department of Physics, Pennsylvania State University, University Park, Pennsylvania 16802, USA}
\author{Saurav Islam}
\email{ski5160@psu.edu}
 \affiliation{Department of Physics, Pennsylvania State University, University Park, Pennsylvania 16802, USA}
\author{Wilson Yanez}
\affiliation{Department of Physics, Pennsylvania State University, University Park, Pennsylvania 16802, USA}
\author{Yongxi Ou}
\affiliation{Department of Physics, Pennsylvania State University, University Park, Pennsylvania 16802, USA}

\author{Haiwen Liu}
\affiliation{Center for Advanced Quantum Studies, Department of Physics, Beijing Normal University, Beijing 100875, China}

\author{X. C. Xie}
\affiliation{International Center for Quantum Materials, School of Physics, Peking University, Beijing 100871, China}
\affiliation{Hefei National Laboratory, Hefei 230088, China}

\author{Juan Chamorro}
\affiliation{Department of Chemistry, Johns Hopkins University, Baltimore, Maryland 21218, USA}
\author{Tyrel M. McQueen}
\affiliation{Department of Chemistry, Johns Hopkins University, Baltimore, Maryland 21218, USA}
\author{Nitin Samarth}
\email{nxs16@psu.edu}
\affiliation{Department of Physics, Pennsylvania State University, University Park, Pennsylvania 16802, USA}

\begin{abstract}

Time-reversal invariance and inversion symmetry are responsible for the topological band structure in Dirac semimetals. These symmetries can be broken by applying an external magnetic or electric field, resulting in fundamental changes to the ground state Hamiltonian and a topological phase transition. We probe these changes via the magnetic-field dependence and gate voltage-dependence of universal conductance fluctuations in top-gated nanowires of the prototypical Dirac semimetal Cd$_3$As$_2$. As the magnetic field is increased beyond the phase-breaking field,we find a factor of $\sqrt{2}$ reduction in the magnitude of the universal conductance fluctuations, in agreement with numerical calculations that study the effect of broken time reversal symmetry in a 3D Dirac semimetal. In contrast, the magnitude of the fluctuations increases monotonically as the chemical potential is gated away from the charge neutrality point. This effect cannot be attributed to broken inversion symmetry, but can be explained by Fermi surface anisotropy. The concurrence between experimental data and theory in our study provides unequivocal evidence that universal conductance fluctuations are the dominant source of intrinsic transport fluctuations in mesoscopic Cd$_3$As$_2$ devices and offers a promising general methodology for probing the effects of broken symmetry in topological quantum materials.

\end{abstract}

\maketitle

%\tableofcontents
\section{Introduction}

The past decade has seen enormous interest in the study of topological band structures created by the interplay between fundamental symmetries and strong spin-orbit coupling in a variety of quantum materials \cite{hasan2010colloquium,Samarth_NMat_2017,yan2017topological,Wen_RevModPhys.89.041004,armitage2018weyl}. Dirac semimetals, a three-dimensional analog of graphene, are an important subset in this materials class, characterized by Dirac states in the bulk with degenerate Weyl nodes that are protected by the presence of both time-reversal symmetry (TRS) and inversion symmetry (IS) \cite{yan2017topological,armitage2018weyl}. The response of Dirac semimetals to applied electrical and magnetic fields has been a matter of active discourse, in bulk crystals\cite{xiong2015evidence,shekhar2015extremely,huang2015observation,liang2015ultrahigh,zhang2016signatures,collins2018electric,ong2021experimental}, thin films \cite{di2021progress,wang2013three,uchida2017quantum,schumann2018observation,kealhofer2020topological,kealhofer2022controlling,Yanez_PhysRevApplied.16.054031,Xiao_PhysRevB.106.L201101}, and patterned micro/nanostructures \cite{li2015giant,moll2016transport,wang2016universal,rashidi_APL_2023}.
An important question in this context is whether one can experimentally observe the expected transformation of a Dirac semimetal into a Weyl semimetal in a given material when the degeneracy of the Weyl nodes is removed by breaking TRS in an external magnetic field. Although angle resolved photoemission spectroscopy (ARPES) could in principle be used to observe such a topological phase transition, it is technically impractical because of the need for a magnetic field. The observation of qualitative changes in the magnetoresistance in a Dirac semimetal at large external magnetic field has provided strongly suggestive evidence for the field-induced transition to a Weyl semimetal in ZrTe$_5$ \cite{Zheng_PhysRevB.96.121401}, but this is still not definitive. We propose that the measurement of universal conductance fluctuations (UCF) potentially provides a more rigorous route to answering this question \cite{hu2017numerical}. 

\begin{figure*}
   %\centering
    \includegraphics[width=16cm]{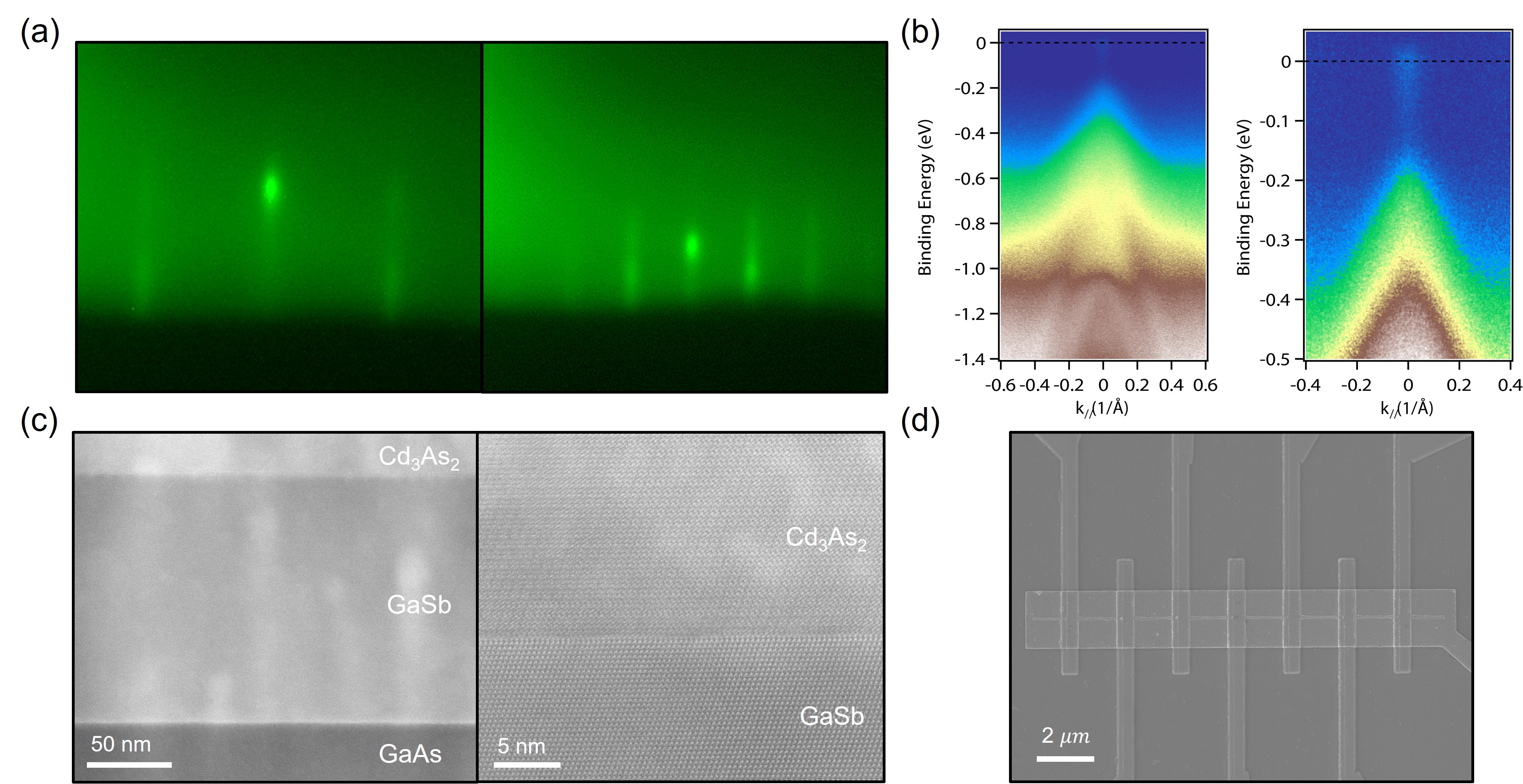}
    \caption{Material growth and device fabrication. (a) RHEED images captured during growth of Cd$_3$As$_2$ films. The electron beam is directed along $[\overline{2}11]$(left) and $[0\overline{1}1]$ (right).(b) ARPES spectra of a 10 nm thick Cd$_3$As$_2$ film grown using similar conditions (substrate temperature, beam flux) as the thicker samples measured in transport. The measurements are taken at $T = 300$ K along the $\bar{K} - \bar{\Gamma}- \bar{K}$ direction. The right panel shows a slightly zoomed in view of the data in the left panel. (c) Cross-sectional HAADF-STEM image of a $20$~nm Cd$_3$As$_2$ film. (d) A scanning electron microscope image of a typical mesoscopic device.} 
    %\vspace{-1.5em}
    \label{1}
\end{figure*}

UCF, a consequence of quantum interference, are aperiodic, reproducible fluctuations of the conductance of magnitude $\approx e^2/h$ in a system, observed when the sample length is comparable to the phase coherence length ($l_\phi$) ~\cite{lee1985universal,feng1986sensitivity,altshuler1986repulsion,Altshuler_1985,Altshuler_Spivak_1985}. The magnitude of UCF is strongly influenced by the underlying symmetries of the system and has been used to probe the ground state symmetries in many materials~\cite{dyson1962statistical,beenakker1997random,ghosh2000universal,mehta2004random,pal2012direct,shamim2014spontaneousSS,islam2018universal,hsieh2021TRS,islam2022benchmarking}. Theory predicts that the magnetic-field-induced topological phase transition from a Dirac semimetal to a Weyl semimetal will manifest as a reduction in UCF magnitude by $\sqrt2$ \cite{hu2017numerical} as one breaks TRS. However, prior experiments have shown an approximate reduction by a factor of $2\sqrt2$ ~\cite{wang2016universal}. Applying an electric field to a Dirac semimetal can also break IS. The effect of this symmetry-breaking perturbation on UCF in a Dirac semimetal is of equal fundamental importance to that of broken TRS but remains unexplored. Here, we address the effect of applying both magnetic and electric fields on UCF in epitaxially-grown thin films of the protypical Dirac semimetal Cd$_3$As$_2$, patterned into nanowires. We observe a factor of $\sqrt{2}$ reduction in UCF amplitude as the TRS is broken with application of a magnetic field, consistent with theoretical predictions. We also observe a monotonic enhancement of the UCF magnitude as the chemical potential is increased using electrostatic gating. We argue that this most likely arises due to Fermi surface anisotropy.

Our experiments provide unambiguous proof of UCF to be the intrinsic source of fluctuations in mesoscopic Cd$_3$As$_2$ devices and further establish its suitability for probing topological phase transitions.  

\begin{figure*}[t]
    %\centering
    \includegraphics[width=16cm]{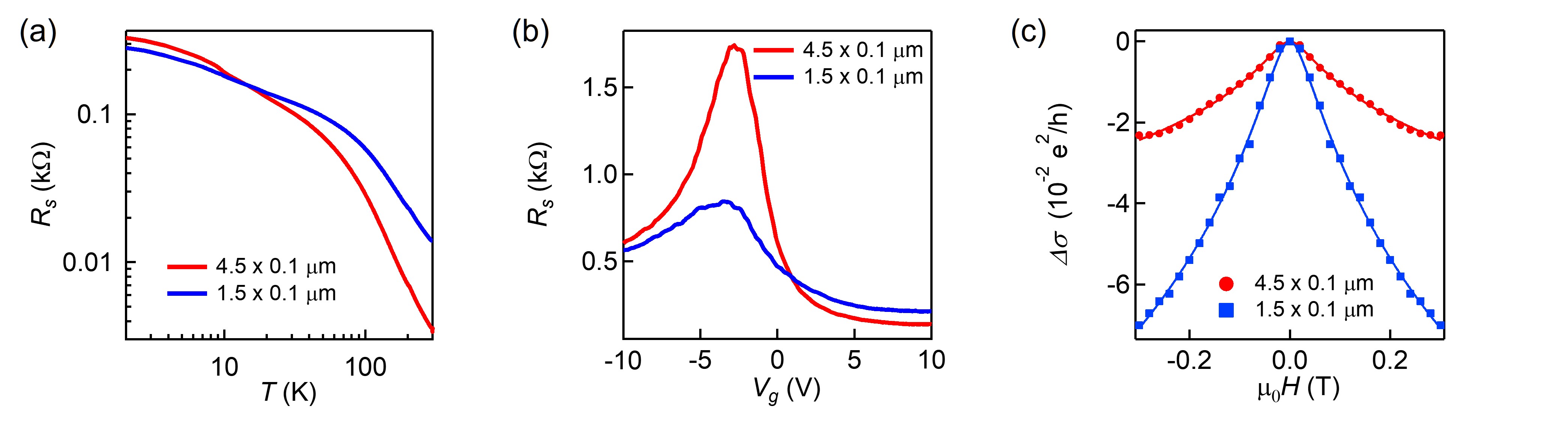}
    \caption{\textbf{Basic electrical characterization :} (a) Sheet resistivity ($R_s$) vs temperature ($T$) of Dev I and Dev II, both showing insulating behavior. (b) Resistance as a function of gate-voltage ($V_g$) of Dev I and Dev II at $T=0.41$~K, exhibiting a charge neutrality point at $V_g=-2.8$~V and $-3.1$~V, indicating the devices are \textit{n}-doped. (c) The quantum correction to conductivity ($\Delta\sigma$) as a function of the magnetic field ($B$) in both devices, exhibiting weak anti-localization at $V_g=0$~V. The blue and red lines show a fit to the data using Eq~\ref{eq:HLN}.}
     \label{2}
 \end{figure*}   
 
\section{Material growth and device fabrication}
Our experiments used Cd$_3$As$_2$ films (20 nm thickness) grown by molecular beam epitaxy (MBE) on semi-insulating (111) GaAs substrates after the deposition of a $100$~nm thick buffer layer of GaSb. During the growth of Cd$_3$As$_2$, we used a high purity compound source of Cd$_3$As$_2$ in a standard effusion cell and a beam equivalent pressure of $1.2\times10^{-7}$ Torr; the substrate temperature was 110~$^{\circ}$C (calibrated using band-edge infrared thermometry). We have established that these growth conditions in our MBE chamber yield Cd$_3$As$_2$ films of good structural and electronic quality oriented in the [112] direction~\cite{Xiao_PhysRevB.106.L201101}. {\it In-situ} reflection high-energy electron diffraction (RHEED) measurements showed streaky patterns in both $[\overline{2}11]$(left) and $[0\overline{1}1]$ directions, indicating reasonably ordered growth~\textcolor{blue}{Fig.~\ref{1}(a)}. The presence of a Dirac cone in MBE-grown Cd$_3$As$_2$ films synthesized under nominally identical conditions is confirmed by {\it in vacuo} ARPES measurements performed after using a vacuum suitcase to transfer films from the MBE chamber to a local measurement chamber. The ARPES measurements were carried out at $T = 300$ K using excitation by the $21$~eV helium I$\alpha$ spectral line from a helium plasma lamp isolated via a monochromator and detection of emitted photo-electrons using a Scienta-Omicron DA 30L analyzer with a spectral resolution of $6$~meV. As shown in \textcolor{blue}{Fig.~\ref{1}(b)}, the ARPES data show the expected linearly dispersing Dirac bands, consistent with previous ARPES measurements from the (112) surface in cleaved bulk samples~\cite{liu2014stable,borisenko2014experimental,yi2014evidence} and thin films \cite{Yanez_PhysRevApplied.16.054031}. We note that calculations and quantum transport measurements\cite{Xiao_PhysRevB.106.L201101} of similar [112]-oriented Cd$_3$As$_2$ thin films indicate that a small quantum confinement-induced gap should be expected at the Dirac point for 20 nm thick films studied here, but our ARPES measurements do not have the resolution to measure this gap. High-resolution transmission electron microscope (TEM) images obtained in cross-section confirm the growth of crystalline Cd$_3$As$_2$ films in the correct phase, as shown in \textcolor{blue}{Fig.~\ref{1}(c)}. 

To fabricate the mesoscopic nanowire devices investigated in this manuscript, we used electron beam lithography to first pattern and deposit $10/30$~nm Cr/Au electrodes using e-beam evaporation. This was followed by another round of lithography and Argon plasma etching to pattern the nanowires. To control the chemical potential, we use a top gate with a $30$~nm {Al$_2$O$_3$} dielectric layer deposited using atomic layer deposition. A scanning electron micrograph of a typical device is shown in \textcolor{blue}{Fig.~\ref{1}(d)}. The transport measurements were performed using a standard four-probe ac technique with a lock-in amplifier in a pumped He-3 Oxford Heliox system. We used a constant current circuit with an excitation current of $10$~nA to reduce Joule heating and a carrier frequency of $17.777$~Hz.

 \section{Electrical Transport Measurements}
 
The temperature dependence of the sheet resistivity ($R_s$) of these films shows insulating behaviour, with $R_s$ increasing monotonically as temperature $T$ is reduced~\cite{cheng2016thickness} (\textcolor{blue}{Fig.~\ref{2}(a)}). %The activated behavior possibly arises due to thermally generated bulk carriers.
$R_s$ as a function of gate-voltage ($V_g$), measured in two channels of length $L=4.5$~$\mu$m (Dev I) and $1.5$~$\mu$m (Dev II),  and width $W=0.1$~$\mu$m, shows maxima $V_g=-2.8$~V and $-3.1$~V respectively, referred to as the charge neutrality point (CNP), indicating the sample to be \textit{n}-doped (\textcolor{blue}{Fig.~\ref{2}(b)}). For the nanowire with $L=4.5$~$\mu$m, the field-effect mobility ($\mu$) of carriers, calculated using $\sigma=ne\mu$ is $8812$~cm$^2$V$^{-1}$s$^{-1}$ and $2475$~cm$^2$V$^{-1}$s$^{-1}$ for the electron and hole channels, respectively. For the nanowire with $L=1.5$~$\mu$m, $\mu$ for electrons and holes are $6375$~cm$^2$V$^{-1}$s$^{-1}$ and $1275$~cm$^2$V$^{-1}$s$^{-1}$, respectively.
The magneto-resistance of both the channels shows weak anti-localization, as expected for a spin-orbit coupled system (\textcolor{blue}{Fig.~\ref{2}c})~\cite{bergmann1984weak,hikami1980spin,chandrasekhar1991weak,mcconville1993weak,wang2011evidence,islamlowt}. In the case of strong spin-orbit coupled systems such as Cd$_3$As$_2$, where $(\tau_{\phi}>>\tau_{so},\tau_{e})$, the quantum correction to conductivity ($\Delta\sigma$) can be fitted with the Hikami-Larkin-Nagaoka (HLN) equation~\cite{hikami1980spin}, given as:
\begin{equation}
\triangle\sigma=\alpha\frac{e^{2}}{\pi h}\left[\psi\left(\frac{1}{2}+\frac{B_{\phi}}{B}\right)-\ln\left(\frac{B_{\phi}}{B}\right)\right]\label{eq:HLN}
\end{equation}
Here, $B_{\phi}$ is the phase coherence field, and $\alpha$ is a fitting parameter. The phase coherence length $l_{\phi}$ can be extracted using $l_{\phi}=\sqrt{\hbar/4eB_{\phi}}$, where $e$ and $\hbar$ are the electronic charge and reduced Planck's constant respectively. Both channels show comparable $l_\phi \approx 200$~nm at $V_g=0$~V.

%$l_{\phi}^{MR}$ in our samples varies from ??? at $V_g=??$  to ???? at high $V_g$'s. 

\begin{figure*}[t]
    %\centering
    \includegraphics[width=16cm]{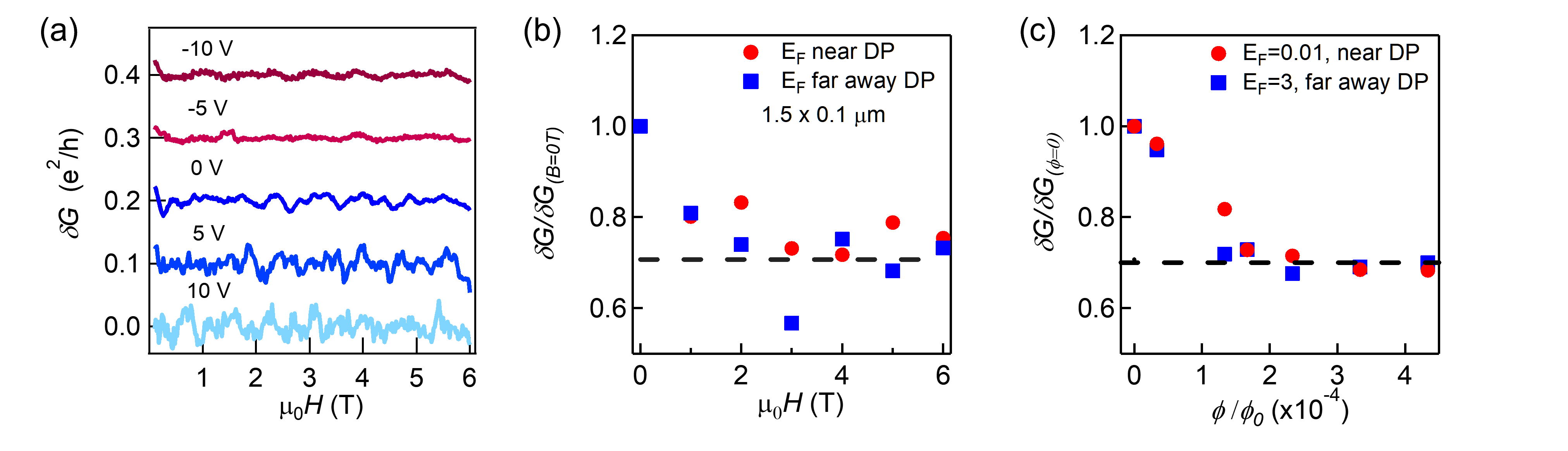}
    \caption{\textbf{Magnetic-field dependence of UCF:} (a) Conductance fluctuations obtained by sweeping the magnetic field at different gate voltages. (b) The magnitude of the fluctuations, normalized with the value at $B=0$~T, shows a reduction by a factor of $\sqrt{2}$ in Dev I. The dashed line corresponds to $\sqrt{2}$ reduction. (c) Calculated UCF magnitude as a function of the normalised magnetic flux $\phi/\phi_0$ when the Fermi energy is both at the Dirac point and away from it. When the magnetic flux goes beyond a critical value, UCF magnitude decreases by $\sqrt{2}$, consistent with the experimental data.}
    %\vspace{-1.5em}
    \label{3}
\end{figure*}

Conductance fluctuations reported in this manuscript were investigated by capturing the four-probe resistance as a function of magnetic field and gate voltage. To probe the effect of breaking TRS, we recorded $R$ by sweeping the magnetic field at different gate voltages, as shown in \textcolor{blue}{Fig.~\ref{3}(a)}. We chose two $V_g$ windows: (a) when $V_g$ is close to the CNP ($-5$~V to $0$~V) and (b) $V_g$  is away from the CNP ($6$~V to $10$~V). The UCF magnitude ($\delta G$) is defined as the rms-magnitude of the fluctuations, calculated after subtracting the background by fitting a polynomial. The fluctuations are aperiodic and reproducible, which are key features of UCF (see Fig.~\ref{S1} in the Appendix for additional data).
As a function of perpendicular magnetic field, we found that the magnitude of UCF is reduced by a factor of $\sqrt{2}$, in both gate-voltage windows, as shown in \textcolor{blue}{Fig.~\ref{3}(b)}. (See also Fig.~\ref{S2} in the Appendix for additional data from a second device.) 

Within the framework of random matrix theory (RMT)
~\cite{dyson1962statistical,beenakker1997random,mehta2004random}, the magnitude of UCF within a phase coherent box is proportional to
\begin{equation}
\langle\delta G^{2}_{\phi}\rangle\propto\left(\frac{e^{2}}{h}\right)^{2}\frac{ks^{2}}{\beta}\label{eq:UCF RMT}
\end{equation}
Here $\beta$, $s$, and $k$ are the Wigner-Dyson parameter (dependent on the universality class of the system), the degeneracy of the system under investigation, and the number of independent eigenmodes of the Hamiltonian respectively. The value of $\beta$ is $1$, $2$, or $4$ for the orthogonal, unitary, and symplectic symmetry classes respectively. The application of a magnetic field removes TRS, splitting the degenerate Dirac points into a pair of Weyl points (in momentum space). This leads to a transition from the Gaussian symplectic class to the unitary symmetry class. In this scenario, $s$ changes from $2$ to $1$ due to the removal of Kramer's degeneracy, while $\beta$ changes from $4$ to $2$. This results in a factor of $\sqrt{2}$ reduction in UCF magnitude as the magnetic field is increased. Microscopically, the self-intersecting Cooperon modes are suppressed as the magnetic field is applied, leading to a reduction of the number of transport modes by a factor of two. In this scenario, the fluctuations arise due to the classical diffuson modes. 

For further validation, we also performed numerical calculations of the UCF magnitude based on a $\boldsymbol{k} \cdot \boldsymbol{p}$ model developed in Ref.~\cite{hu2017numerical}. 
We adopt the following minimal 4$\times$4 Hamiltonian $H_0$ to describe the 3-dimensional Dirac semimetal $\text{Cd}_{\text{3}}\text{As}_{\text{2}}$
\begin{equation}
    H_0(\textbf{k}) = \left(
   \begin{array}{cccc}
    M(\textbf{k}) & A\textbf{k}_+ &  D\textbf{k}_- & 0 \\
    A\textbf{k}_- & -M(\textbf{k}) &  0 & 0 \\
    D\textbf{k}_+ & 0 &M(\textbf{k}) & -A\textbf{k}_- \\
    0  & 0 &  -A\textbf{k}_+ &  -M(\textbf{k})
   \end{array}
  \right),
\end{equation}\label{EffectiveModel}
with $M(\textbf{k})=M_0-M_zk_{z}^{2}-M_xk_{x}^{2}-M_yk_{y}^{2}$ and $k_{\pm}=k_x\pm ik_y$. $A$ and $D$ are the strength of spin-orbital coupling between the inverted bands $\pm M(\textbf{k})$ and the two $M(\textbf{k})$ orbits, respectively. A real space version of $H_0(\textbf{r})$ on a discretized lattice, $H_0(\textbf{k})$, is obtained through Fourier transformation. The externally applied magnetic field $\vec{B}$ enters $H_0$ through the Piers substitution, e.g., for magnetic field applied along z direction, $t_x\rightarrow t_xe^{i\phi}$, where $\phi$ measures the magnetic flux through a unit lattice square. The disordered $\text{Cd}_{\text{3}}\text{As}_{\text{2}}$ material is modeled using $H=H_0+U(\textbf{r})$, where $U(\textbf{r})$ is an onsite random potential uniformly distributed on $[-W,W]$. 

We numerically compute the zero-temperature conductance $G=\frac{e^2}{h}Tr[\Gamma_LG^r\Gamma_RG^a]$ using the Landauer-B\"{u}ttiker formula~\cite{datta_1995}, where $G^r(E_F)=[G^a(E_F)]^\dag=[E_F-H-\Sigma_L-\Sigma_R]^{-1}$ is the retarded Green's function, $\Gamma_{L/R}=i[\Sigma^r_{L/R}-\Sigma^a_{L/R}]$ is the line width function, and $\Sigma_{L/R}$ is the self-energy of the left/right lead. The conductance fluctuation $\Delta G$ is calculated as the standard deviation of conductance $G$ for an ensemble of disorder, $\Delta G=\langle(G-\overline{G})^2\rangle^{\frac{1}{2}}$, averaged over at least 200 ensembles. In the calculation, we use $M_0=-0.4, M_z=M_x=M_y=-0.5$, and $A=D=1$, and use a quasi-1-dimensional system with sizes $L_x=10, L_y=30, L_z=100$. The magnitude of UCF is determined as the average value over the plateau where conductance fluctuations saturate as disorder strength is varied. We further confirm the convergence of the UCF magnitude by testing its sensitivity to the system size $L_\alpha$, $\alpha=x,y,z$. This method ensures that the UCF obtained are universal values in the diffusive regime and also helps get rid of the finite size effect in numerical calculations.

As shown in \textcolor{blue}{Fig.~\ref{3}c}, when the magnetic flux per unit cell, normalized with the magnetic flux quanta ($\phi/\phi_0$) goes beyond a critical value, the UCF magnitude decreases by $\sqrt{2}$, both close to and away from the charge neutrality point. Thus, our observed reduction in the magnitude of UCF by a factor of $\sqrt{2}$ is consistent with that predicted theoretically for Dirac materials. It is important to emphasize that although the parameters used in the model are not realistic for the experiments, it is sufficient to capture the correct transitions of the intrinsic UCF magnitude between symmetry classes, since the transitions between UCF magnitudes depend only on the symmetry indices that are not dependent on the details of the Hamiltonian parameters, such as the strength of spin-orbital coupling and the effective band mass, etc. However, our theory is unable to capture the critical magnetic field beyond which the UCF changes its intrinsic magnitude as it depends on various system parameters other than the symmetry indices. We also note that previous experiments on mesoscopic Cd$_3$As$_2$ channels showed a $2\sqrt{2}$ reduction~\cite{wang2016universal}. This may be caused by factors such as magnetic-field induced gap or decoherence~\cite{hu2017numerical}. Quantum confinement, which can open up a small gap at the Dirac point in films of the thickness we used ($\approx 20$~nm), may also play a role here since the thickness of our films is much smaller  than the nanowires investigated in the previous experiment ($\approx 100$~nm). The role of disorder is also not clear which leads to the removal of valley degeneracy and can affect the UCF magnitude.

\begin{figure}
    %\centering
    \includegraphics[width=8cm]{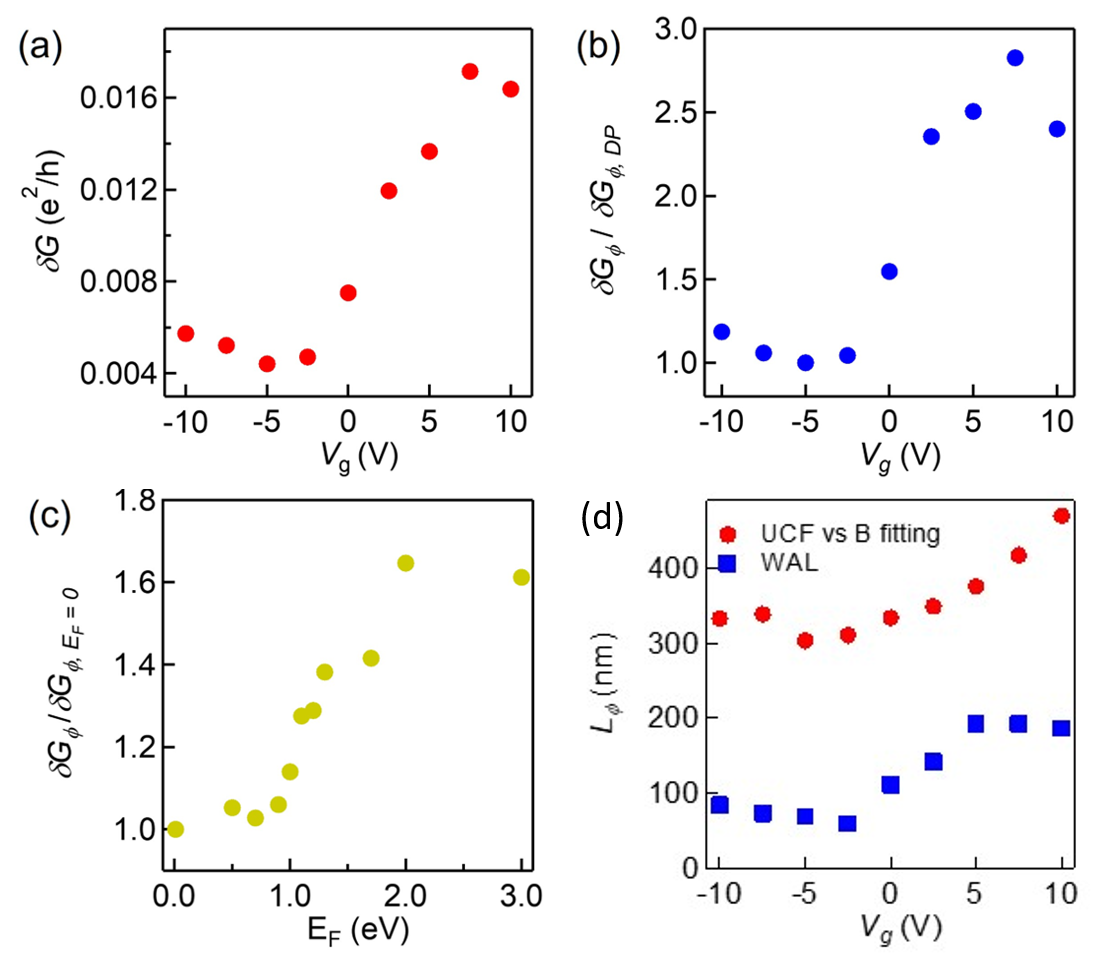}
    \caption{Gate-voltage dependence of UCF. (a) RMS-magnitude of the fluctuations as a function of gate-voltage ($V_G$) (b) UCF magnitude within a phase coherent box, showing an increase by a factor of $2.5$ as $V_g$ tuned away from the charge neutrality point. (c) Normalized UCF magnitude within a phase coherent box as a function of the Fermi energy.(d) Comparison of the phase breaking length ($l_\phi$) obtained from different methods.} 
    %\vspace{-1.5em}
    \label{4}
\end{figure}

To evaluate the impact of an externally applied electric field on UCF, we extracted the rms-magnitude as a function of $V_g$, which is plotted in \textcolor{blue}{Fig.~\ref{4}(a)}. We observe a strong suppression of UCF near the charge neutrality point. The suppression of UCF at the charge neutrality point was observed in prior studies in topological insulators and Dirac semimetals. This was attributed to an increase in $l_\phi$ at high carrier densities due to the screening of impurity scattering potential~\cite{rossi2012universal,wang2016universal,islam2018universal}, although the trend within a phase coherent box was not investigated. In the context of single-layer graphene, although $\delta G$ shows a decrease or increase at higher Fermi energy ($E_f$), the magnitude within a phase coherent box ($\delta G_\phi$) reduces by a factor of four away from the Dirac point, due to the removal of valley degeneracy~\cite{vidyaPRL}. To probe this further, we evaluated the UCF magnitude within a phase coherent box by using $\delta G^2_{\phi}={LW}\frac{\delta G^2}{L^2_{\phi}}$~\cite{pal2012direct}. We observe an increase in the magnitude (normalized with the value at the Dirac point, $\delta G_{\phi,DP}$) as $V_g$ is tuned away from the Dirac point (\textcolor{blue}{Fig.~\ref{4}(b)}), by a factor of $2.5$ in this device (See Appendix B).  This trend of increasing magnitude of UCF with increasing $E_f$ is also captured in our numerical calculation, as shown in \textcolor{blue}{Fig.~\ref{4}c}, where we plot in the normalized UCF magnitude as a function of  $E_f$. We attribute this increase of UCF to Fermi surface anisotropy. The strong anisotropy of the Fermi surface around the Dirac point is an important feature of $\text{Cd}_{\text{3}}\text{As}_{\text{2}}$ and can be described correctly by the effective low-energy Hamiltonian $H_0(\textbf{k})$ in Eq.~\ref{EffectiveModel}~\cite{liu2014stable,zhao2015anisotropic}. The effects of Fermi surface anisotropy can be intuitively understood as follows: we neglect the $D\textbf{k}_{\pm}$ terms in $H_0(\textbf{k})$ and locate the Dirac point at $(0,0,\sqrt{M_0/M_z})$. The states around the Dirac point have much longer wavelength $\lambda_{\alpha}=\frac{2\pi}{k_\alpha}$ along the $\alpha=x,y$ direction than the $z$ direction. Thus the electrons effectively travel in a 1-dimensional system rather than in a 3-dimensional system. UCF are suppressed through the reduction of the prefactor from $c_{d=3}$ to $c_{d=1}$ in the formula $\delta G = c_d \sqrt{\frac{ks^2}{\beta}}$. Away from the Dirac point, electrons recover the 3-dimensional transport and the UCF increase due to the increase of the prefactor $c_{d}$. This is validated in the numerical calculation, where we have used a cube of size $L_x=L_y=L_z=12$ to describe a phase-coherent region inside the material, which has been probed experimentally here. We emphasize here that the effect of Fermi surface anisotropy depends on realistic parameters and can only be described qualitatively through the model assumed in this manuscript; a quantitative understanding of the influence of the Fermi-surface anisotropy on the UCF still needs further study. Another possibility is the change in values of $k$ and $s$. Due to the electric field introduced by $V_g$, the energy band of Cd$_3$As$_2$ around one Weyl point splits into two bands ($k=2$) while still satisfying time-reversal symmetry ($s=2$). Thus, at higher carrier densities, the UCF of quasi-particles are characterized by $k=2$ and $s=2$, while close to the Dirac point, $k=1$ and $s=1$ take place due to the pronounced charge impurity scattering ~\cite{RevModPhys.83.407}. Hence, from Eq.~\ref{eq:UCF RMT}, we get a factor of $2\sqrt{2}$ increase in UCF magnitude at higher $E_f$. The disagreement with the experimental observations is possibly due to the Fermi surface anisotropy.  

Finally, we extracted the $V_{G}$-dependence of $l_{\phi}$
from two independent methods: (a) we determine via the magnetoresistance by using fits to the HLN equation (Eq.~\ref{eq:HLN}); (b) we directly determine $l_{\phi}$ from the UCF by analyzing the auto-correlation function~\cite{vila2007universal,wang2016universal}: 

\begin{equation*}
F(\Delta B)= \frac{\langle\delta G(B) \delta G(B+\Delta B)\rangle_B}{\langle \delta G^2 \rangle} .
\end{equation*}
We use this to obtain the correlation field $B_0$ using the equation $F(B_0)=0.5F(0)$ and then determine $l_\phi=2.4({\frac{h}{eB_0}})^\frac{1}{2}$.

We find that the values of $l_{\phi}$ determined from magnetoresistance and UCF differ in magnitude by a factor of three over the $V_g$ range that has been investigated (\textcolor{blue}{Fig.~\ref{4}((d))}). We also find $l_{\phi}$ increases away from the Dirac point in both cases. Enhanced screening of electromagnetic fluctuations at higher number densities leads to a larger $l_{\phi}$ away from the Dirac point, while inhomogeneity from electron-hole puddles leads to lower $l_\phi$ around the Dirac point ~\cite{chiu2013weak,tian2014quantum,islam2018universal}. The factor of three difference in $l_{\phi}$ obtained from the two methods has two possible explanations. First, the phase breaking time $\tau_{\phi}$ relevant for weak localization (WL)
is related to the Nyquist dephasing rate, while the scattering time scale for UCF depends on the out-scattering time which is related to the inverse of the inelastic collision frequency~\cite{altshuler1982effects,blanter1996electron}. The influence of these two mechanisms on dephasing can differ and this difference has been investigated in a variety of samples~\cite{chandrasekhar1991weak,mcconville1993weak,mcconville1999,hoadley1999experimental,aleiner2002inelastic,trionfi2004electronic,trionfi_2,shamim2017dephasing,islam2018universal}. 
%The other possible reason that can lead to a difference in $l_{\phi}$ obtained from the two methods is that Eq.~{\ref{eq:ucf2}} is strictly only valid in the case of $l_{\phi}<<L,l_{T}$\ \cite{adroguer2012diffusion,akkermans2007mesoscopic}, where $l_{T}$ is the thermal length.
Further investigation is required to identify if both WL and UCF are governed by the same scattering rates in Dirac materials.

\section{Conclusion}

In conclusion, we have investigated UCF in mesoscopic transport channels of the Dirac semimetal Cd$_3$As$_2$. We find that the UCF magnitude is reduced by a factor of $\sqrt{2}$ as the magnetic field is increased due to the removal of time reversal symmetry. Our observations are consistent with a topological phase transition from the symplectic to unitary symmetry class. We also find that the magnitude of UCF increases as the Fermi energy in the system is increased, which we attribute to Fermi surface anisotropy rather than broken inversion symmetry. Our experiments establish UCF to be the intrinsic source of fluctuations in these systems and emphasize their importance in probing phase coherent transport. The good concurrence between theoretical predictions and experimental observations indicates that measurements of UCF provide a promising route for rigorously probing the influence of broken symmetry on the band structure of topological quantum materials. 

We thank Yayun Hu for invaluable technical contributions to the theoretical calculations in this paper. This project was supported by the Institute
for Quantum Matter under DOE EFRC Grant No. DESC0019331.
The Penn State Two-Dimensional Crystal Consortium Materials Innovation Platform (2DCC-MIP) under NSF Grant No. DMR-2039351provided support for ARPES measurements.

%\bibliography{ref.bib}% Produces the bibliography via BibTeX.

\appendix

\section{\label{sec:level1}Key features of UCF. }

\begin{figure}[H]
    \centering
    \includegraphics[width=9cm]{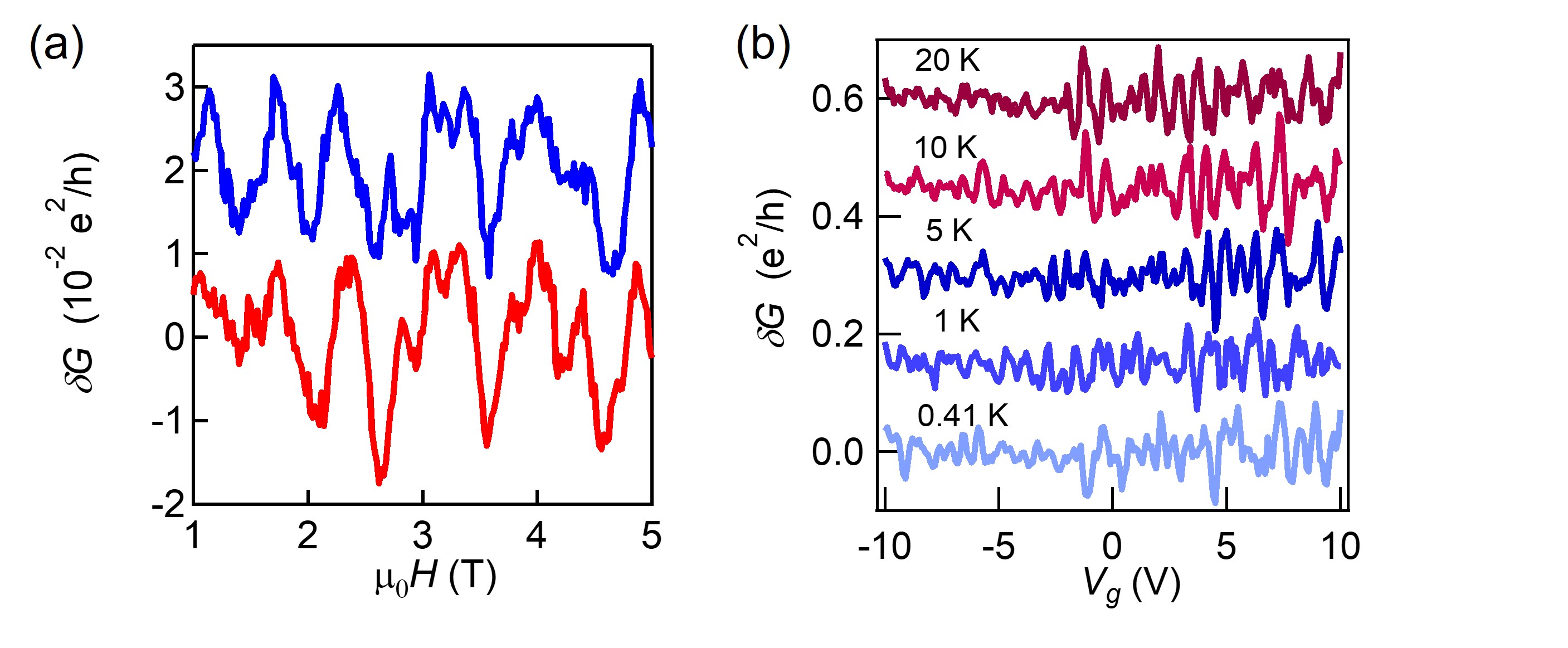}
    \caption{\textbf{Features of universal conductance fluctuations}: (a) Run-to-run reproducibility of the UCF measured at $T=0.41$~K at $V_g=0$~V in Dev I. (b) UCF at different temperatures, showing a reduction in amplitude as the temperature ($T$) is increased. The curves have been offset vertically for clarity.} 
    %\vspace{-1.5em}
    \label{S1}
\end{figure}

\textcolor{blue}{Figure~\ref{S1}(a)} shows the run-to-run reproducibility of the fluctuations as well as their aperiodic nature, two of the key signatures of UCF. The amplitude of the fluctuations also reduces with increasing $T$ as shown in \textcolor{blue}{Fig.~\ref{S1}(b)}, as the interference effect gets suppressed due to thermal fluctuations, another key feature of UCF.

\section{Magnetic field and gate-voltage dependence of UCF in addditional device (Dev II)}
\begin{figure}[H]
    \centering
    \includegraphics[width=9cm]{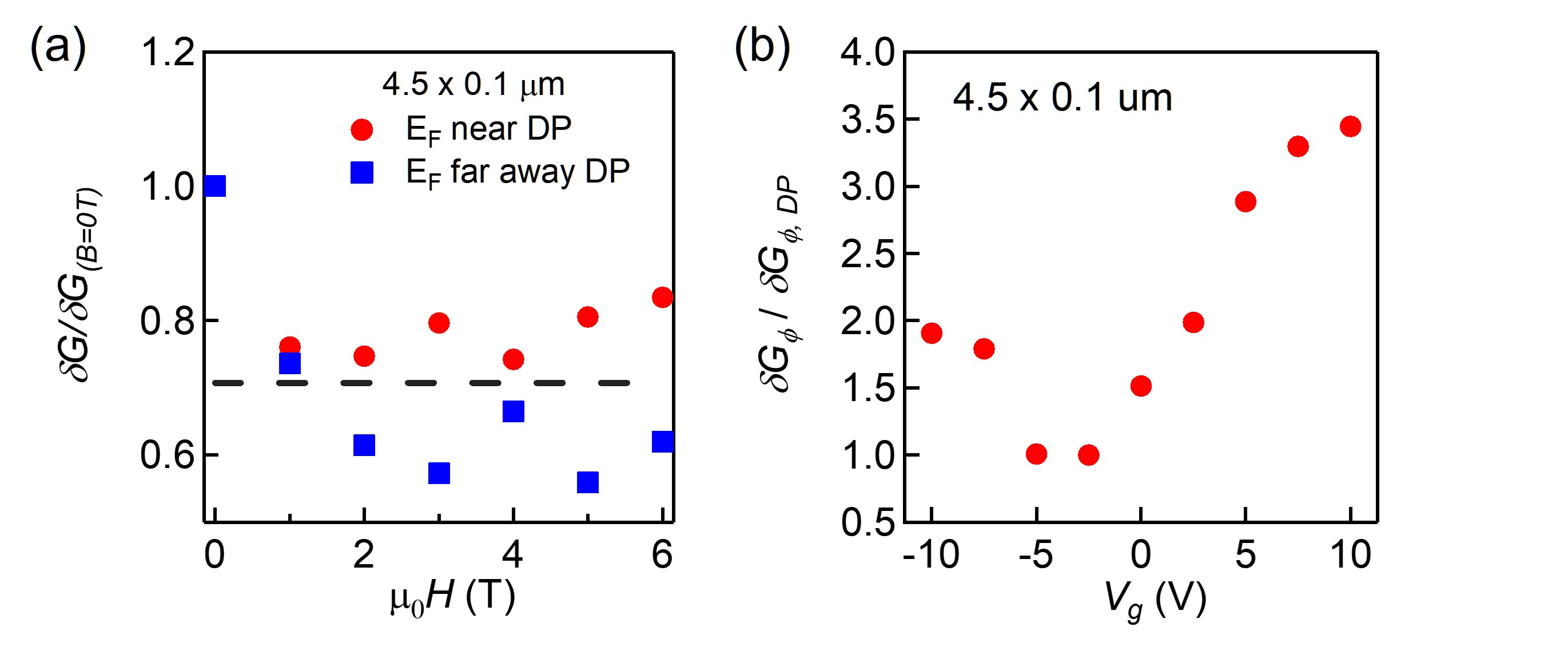}
    \caption{\textbf{Magnetic-field and gate-voltage dependence of UCF in Dev II}:(a) The magnitude of the fluctuation, normalized with the value at B = 0 T, shows a reduction by a factor of $\sqrt{2}$ in Dev II.(b) UCF magnitude within a phase coherence box, showing an increase as $V-g$ is tuned away from the charge neutral point.} 
    %\vspace{-1.5em}
    \label{S2}
    \end{figure}

\textcolor{blue}{Figures~\ref{S2}(a)} and \textcolor{blue}{\ref{S2}(b)} show the UCF for Dev II as a function of magnetic field and gate-voltage. The behavior of UCF in Dev II is similar to that observed in Dev I. 

\providecommand{\noopsort}[1]{}\providecommand{\singleletter}[1]{#1}%

\end{document}